\documentclass[nofootinbib,onecolumn,preprintnumbers,amsmath,amssymb,prd]{revtex4}

\preprint{IPM/P-2009/051}
\allowdisplaybreaks

\begin{document}
\title{Visiting Horava-Lifshitz gravity in extra dimensions}

\author{Qasem \surname{Exirifard}}
\noaffiliation 
\email{exirifard@gmail.com}

\begin{abstract}
The Horava-Lifshitz gravity, having broken the symmetry of space and time, includes three objects: the spatial metric $g_{ij}$, the lapse variable $N$, and the shift variable $N_{i}$. Each of these objects have their own scaling dimensions. The action of the theory is required to be invariant under this scaling, and spatial diffeomorphism, and  the temporal foliation symmetry. Noting that action can non-trivially  depend  on the lapse variable, we suggest to consider the Horva's approach to quantum gravity in higher dimensions such that a set of extra spacial coordinates  possess a scaling dimension different from that of rest. In so so doing, we propose a  new power counting renormalizable theory   for quantum gravity in its UV point. We show that the IR point and UV point of the proposal possess the same number of degrees of freedom in $3$, $8$ and $27$ extra-dimensional space-time geometry.   
\end{abstract}

\maketitle
From the far east to the far west, once for a long time, all the known great minds of the time unanimously agreed on the first theory of every thing: the doctrine of four elements, a logical extension of the one element (water) theory of Babylonia and Assyria. Today we have transcended/left this doctrine, and each of us keeps smiling at that theory of every thing. But the moment we halt smiling we find ourselves standing on nothing but the empty void of between the two peaks of understandings. One peak is the theory of general relativity, and the other one is quantum field theory. Even at their very basic of their foundations, these two peaks of knowledge contradict the nature of each other. At the top of the peaks, this contradiction is manifested by our total lack of ability to quantise general relativity. The weakest form of how to quantise gravity is to find a path from one peak to the other. Unfortunately, so far, we have not yet found any connecting path between these two that can be argued to be reasonably free of obstacles.

Petr Horava recently has suggested, to move away from one peak, and to break the symmetry between time and space in the UV point in order to have a power-counting renormalizable theory at that point. In this short note we would like to provide a deconstruction of the Horava's proposal which retain time and most of the space symmetric, but makes the space asymmetric. The deconstruction is implemented first by reviewing the terms that Petr Horava has missed to include in his proposal, then using these missing terms to  create the deconstruction.

\section{Revising the Horava's proposal}
Ref. \cite{Horava} has considered only functionals of the metric when possible forms for the potential term were being investigated. However, it addition to the metric, the theory includes the lapse and shift variable. The potential could be a non-trivial functional of these variables. We first show/review the non-trivial dependence on the lapse variable. 

Let us start with a $D+1$ dimensional metric of the space-time
\begin{equation}
ds^2= g_{\mu\nu} dx^\mu dx^\nu\,. 
\end{equation}
Using the coordinate $(t,x^i)$, the metric can be re-written in the ADM form:
\begin{equation}
ds^2 = - N^2 dt^2 + g_{ij} (dx^i + N^i dt) (dx^j+N^j dt)\,. 
\end{equation}
Ref. \cite{Horava} assigns the following scaling dimensions to the coordinates and the variables appeared in the ADM form of the metric:
\begin{equation}
[t]=-z,\, [x]=-1,\, [N]=0,\, [N_i]=z-1,\, [g_{ij}]=0\,. 
\end{equation}
It then requires the physical quantities to be invariant under the scaling symmetry. Furthermore ref.\cite{Horava} demands the physical quantities to be invariant under the following coordinate transformations (foliation symmetry):
\begin{eqnarray}
 t &\to & f(t)\,,\\
 x^i &\to & \xi^{i}(x^j,t)\,. 
\end{eqnarray}
Under $x^i \to  \xi^{i}$, $N$ transforms as a scalar. Under $t \to f(t)$,  $N$ transforms as a scalar density. We can construct the following quantify out from $N$  which transforms covariantly (as a tensor) under both of these transformations:
\begin{eqnarray}
\nabla^i \ln N\,,
\end{eqnarray}
The scaling dimension of the above tensor read
\begin{equation}
[\nabla^i \ln N] = 1\,,
\end{equation}
Ref. \cite{Horava} first constructs the  kinetic term for $g_{ij}$. It then adds the potential term for the metric to the kinetic term. The potential term can be any scalar respecting the symmetries of the theory constructed out of the $g_{ij}$, $R_{ijkl}[g_{ij}]$, $N_i$ and $N$ and their spacial covariant derivatives;
\begin{equation}
S_V = S_V[g_{ij}, R_{ijkl}, N_i, N, \nabla_i]\,, 
\end{equation}
A subclass of which reads
\begin{equation}
 S_V = \int dt d^Dx \sqrt{\det g} N V[g_{ij}, R_{ijkl}, N_i, N, \nabla_i] 
\end{equation}
where $V[g_{ij}, R_{ijkl}, N_i, N, \nabla_i]$ is a scalar under spacial transformation and preserves the foliation symmetry. Ref.\cite{Horava} from outset presumes that $V[g_{ij}, R_{ijkl}, N_i, N, \nabla_i]$ does not depend on the lapse and shift variables. However in contrary to the assumption of the ref.\cite{Horava}, the potential can non-trivially depend on the lapse  variable and its covariant derivatives. In fact any spacial scalar constructed out of $g_{ij}$, $g^{ij}$, $R_{ijkl}[g_{ij}]$, $\nabla_{i} \ln N$ and their spatial covariant derivatives will preserve the foliation symmetry. So the `general' potential can be written as:
\begin{eqnarray}
S_V = \int dt d^Dx \sqrt{g}\, N\, V\,,\\
V = V[g_{ij}, R_{ijkl}, \nabla_i \ln N, \nabla_i]\,, 
\end{eqnarray}
where $V$ is a scalar constructed out from its arguments. Because $S_V$ should be of vanishing scaling dimension, $V$ must satisfy $[V]= D+z$. For sake of simplicity we consider only the choice of $D=3$ and $z=3$  but our conclusion apparently remains valid for other cases of interest. For this choice we have
\begin{equation}
[\nabla_i]=1, [\nabla_i N]=1, [g_{ij}]=0\,. 
\end{equation}
Let $f^{i_1\cdots i_n}=f^{i_1\cdots i_n}[\nabla^i \ln N, \nabla^i]$ be defined as the most general polynomial constructed out from $\nabla \ln N$ and $\nabla$ that carries the ``$i_1\cdots i_n$'' indices:
\begin{eqnarray}
f^{i_1} &=& c_1 \nabla^{i_1} \ln N\,,\\ 
f^{i_1 i_2} &=& c_1 \nabla^{i_1} \ln N \nabla^{i_2} \ln N + c_2 \nabla^{i_1} \nabla^{i_2} \ln N\,,\\
f^{i_1 i_2 i_3} &=& c_1 \nabla^{i_1} \ln N \nabla^{i_2} \ln N \nabla^{i_3} \ln N+ c_2 \nabla^{i_1} \nabla^{i_2} \ln N\,\nabla^{i_3} \ln N + c_3  \nabla^{i_1} \nabla^{i_2}\nabla^{i_2} \ln N\,,\\
f^{i_1 i_2 i_3 i_4} &=& c_1 \nabla^{i_1} \ln N \nabla^{i_2} \ln N \nabla^{i_3} \ln N \nabla^{i_4} \ln N+ c_2 \nabla^{i_1} \nabla^{i_2} \ln N\,\nabla^{i_3} \ln N \nabla^{i_4} \ln N+\nonumber\\
&&+ c_3  \nabla^{i_1} \nabla^{i_2}\nabla^{i_2} \ln N\, \nabla_{i_4} \ln N  + c_4 \nabla^{i_1} \nabla^{i_2}\nabla^{i_2} \nabla^{i_4} \ln N + c_5 \nabla^{i_1} \nabla^{i_2} \ln N\,\nabla^{i_3} \nabla^{i_4} \ln N,
\end{eqnarray}
where $c_i$'s  are real constant parameters, and the explicit form of $f^{i_1\cdots i_5}$ and $f^{i_1\cdots i_6}$ have not been written but they can be induced. Using the above definition it follows that $V$ reads
\begin{equation}\label{Vg}
V = \sum_{n=0}^6 V^{6-n}_{i_1\cdots i_n}[g]\, f^{i_1\cdots i_n}[\nabla^i \ln N, \nabla^i]\,,
\end{equation}
wherein $V^{6-n}_{i_1\cdots i_n}[g]$ is a tensor of rank $n$ containing $6-n$ derivatives constructed out from the  spatial metric, and its spatial derivatives. For $D=3$ these terms can  easily be constructed because the Ricci tensor carries all the degrees of freedom of the Riemann tensor. We suffice to present  $V^{5}_{i_1}[g]$ and $V^{4}_{i_1 i_2}[g]$
\begin{eqnarray}
 V^{5}_{i_1}[g] &=&  d_1 \nabla_{i_1} \Box R + d_2 R \nabla_{i_1} R\, + d_3  R_{\alpha\beta} \nabla_{i_1} R^{\alpha\beta} + d_4 R_{i_1 \beta} \nabla^\beta R , \\
V^{4}_{i_1 i_2}[g] &=& d_1 R_{i_1 i_2} R + d_2 R_{i_1 \alpha} R^{\alpha}_{~i_2} + d_3 R^2\, g_{i_1 i_2} + d_4 \nabla_{i_1}\nabla_{i_2} R +d_5 \nabla_{i_2}\nabla_{i_1} R + d_6 \Box R_{i_1 i_2} + d_7 \Box R\, g_{i_1 i_2}
\end{eqnarray}
where $R$ stands for the Ricci scalar and $d_i$'s are real constant parameters.  In higher dimensions one may use the classification of ref. \cite{Fulling} to write the algebraically independent terms in $V^{6-n}_{i_1\cdots i_n}[g]$. Note that some of the above terms are related to each other by partial integration in the action. We leave implementing the simplification by partial integration to interested reader. These terms has been proposed in \cite{Blas:2009yd} and the physical significance to resolve the problems like that of highlighted in \cite{Charmousis:2009tc,Bogdanos:2009uj,Bogdanos:2009uj} is investigated in \cite{Blas:2009qj}, some claims of which are yet being debated \cite{Li:2009bg,Blas:2009ck}. In the following we aim to use these missing terms in order to propose a different Lifshitz point  serving as a candidate for the UV completion of Einstein general relativity.

To this aim we recall that, so far, different scaling dimensions have been assigned to time and space in order to make the theory power counting renormalizable. Let's assume that there exists at least one quantum-mechanically effective dimension beside the ordinary space-time dimensions. We represents this dimension as $q$. Using coordinate $(q,x^\mu)$, the effective metric in the  ADM-like coordinates reads
\begin{equation}
ds^2 \, = \,  N^2 dq^2 + g_{\mu\nu} (dx^\mu + N^\mu dq) (dx^\nu + N^\nu dq)   
\end{equation}
where $\mu$ runs from $0$ to $D$. From this time on, $D$ will represent the dimension of the ordinary space-time. As before, we can assign the following scaling coordinates to the variables appeared above
\begin{equation}
[q]=-z,\, [x]=[t]=-1,\, [N]=0,\, [N_\mu]=z-1,\, [g_{\mu\nu}]=0\,. 
\end{equation}
In the same analogy, we define the foliation symmetry over the $q$ coordinate:
\begin{eqnarray}
 q &\to & f(q)\,,\\
 x^\mu &\to & \xi^{\mu}(x^\nu,q)\,. 
\end{eqnarray}
We then require the theory to be invariant under scaling and foliation of the $q$ coordinate. This theory can be made power counting renormalizable. Having defined 
\begin{equation}
K_{\mu\nu} \,=\, \frac{1}{2 N} (\frac{\partial g_{\mu\nu}}{\partial q}-\nabla_\mu N_\nu - \nabla_\nu N_\mu) 
\end{equation}
the part of the action of the theory that includes only two derivatives follows
\begin{equation}\label{sk}
S_k \, = \, \frac{2}{k^2} \int dq d^D x \sqrt{-\det g} N (K_{\mu\nu} K^{\mu\nu}-\lambda K^2)\,.  
\end{equation}
In contrary to the original Horava-Lifshitz theory, this part does not include the space-time derivatives of the metric. We set $z=D$ and make the coupling constant for this action dimensionless.  In our realisation, the term that is known as potential term of the Horava theory generates the dynamics of the metric. That term reads
\begin{equation}
S_V \,=\, \int dq d^{D} x \sqrt{-\det g} N V[g_{\mu\nu},\nabla_\mu \ln N,\nabla_\mu] 
\end{equation}
while 
\begin{equation}\label{Vgm}
V = \sum_{n=0}^{2D} V^{2D-n}_{i_1\cdots i_n}[g]\, f^{i_1\cdots i_n}[\nabla^i \ln N, \nabla^i]\,,
\end{equation}
wherein $V^{2D-n}_{i_1\cdots i_n}[g]$ is a tensor of rank $n$ containing $2D-n$ derivatives constructed out from the  space-time metric, and its covariant derivatives. $f^{i_1\cdots i_n}$ again is the most general polynomial constructed out from $\nabla \ln N$ and $\nabla$ that carries the ``$i_1\cdots i_n$'' indices. Let $V$ be linear in term of the Riemann tensor. In other words, consider just $n=2D-2$ term in \eqref{Vgm}. Let $f^{i_1\cdots i_{2D-2}}$  be
\begin{equation}
f^{i_1\cdots i_{2D-2}} \,=\, \prod_{p=1}^{2D-2} \nabla^{i_p} \ln N. 
\end{equation}
These choices lead to the following `potential' term:
\begin{equation}
V = V^{2}_{i_1\cdots i_{2D-2}}[g]  \prod_{p=1}^{2D-2} \nabla^{i_p} \ln N.
\end{equation}
the first variation of which leads to  third order differential equations of motion for $g$ and $N$. Usually it is the appearance of four order derivatives  and beyond that leads to fluctuations which are hard to control or to be considered consistent with causality. Third order equations should not be discarded based on the ghost-free criterion from outset. Besides there exist some possibilities in $V = V^{2}_{i_1\cdots i_{2D-2}}[g]$ to further simplify or  demand additional properties on the equations of motion.

To summarize, 
\begin{equation}
S =  \frac{2}{k^2}\int dq d^D x \sqrt{-\det g} N (K_{\mu\nu} K^{\mu\nu}-\lambda K^2 + V^{2}_{i_1\cdots i_{2D-2}}[g]  \prod_{p=1}^{2D-2} \nabla^{i_p} \ln N)
\end{equation}
is a power counting renormalizable theory that possesses all symmetries of a $D$ dimensional space-time (enhanced by foliation symmetry over the $q$ coordinate). So it can serves as a candidate for UV fixed point of gravity in a $D$ dimensional space-time. In contrast to the original Horava's proposal, it does not break the symmetry between time and space. Instead it breaks the symmetry in space, it divides the space into the $q$ coordinate and the ordinary directions. Its low energy limit is gravity in $D+1$ dimensional space-time where the symmetry group for the spacial directions is broken.

Since in the low energy limit the lowest derivative terms are the leading terms, one expects to have the usual Einstein-Hilbert action in $D+1$ dimension corrected by $S_k$ \eqref{sk}
\begin{equation}
S_{\text{Low-Energy}} \,=\, \int d^Dx dq \sqrt{-\det \tilde{g}} \tilde{R} + S_k\,, 
\end{equation}
where $\tilde{g}$ and $\tilde{R}$ are the metric and the Ricci scalar in/of the higher dimensional space-time. Note that $S_k$ manifestly breaks the symmetry between the ordinary space and the $q$ coordinate.  The phenomenology of original Horava's proposal has investigated to almost fine details, for example \cite{Saridakis:2009bv,Cai:2009in,Leon:2009rc,Dutta:2009jn}. There exists a fair amount of literature/models about phenomenology of extra dimensions. In each of these models, it sounds interesting how the inclusion of $S_k$ term  affects the consistency of the model with cosmological observation, and possibly alters its prediction.  

Also note that we can increase the number of extra dimensions from one to an arbitrary number, call it $n$, where the metric reads
\begin{equation}\label{extra-q}
ds^2 \,=\, N_{ab} dq^a dq^b + g_{\mu\nu}(dx^\mu + N^\mu_a dq^a )(dx^\nu + N^\nu_a dq^a )
\end{equation}
where $\mu$ denotes the ordinary $D$ dimensional space-time, and $a$ runs from $1$ to $n$, and the scaling dimension of the variables read
\begin{equation}
[q_a]=-z,\, [x]=[t]=-1,\, [N_{ab}]=0,\, [N^\mu_a]=z-1,\, [g_{\mu\nu}]=0\,. 
\end{equation}
Note that we shall use $N_{ab}$ to brings down/up the indices of $q$ dimensions. The generalised foliation symmetry on q-dimensions reads
\begin{eqnarray}\label{gen-foil}
q_n &\to& f_n(q_m)\\\nonumber
x_\mu &\to& \xi_\mu(x_\nu,q_n)
\end{eqnarray}
The `potential' term follows
\begin{equation}\label{Vgen}
S_V \, = \, \int d^nq d^D x \sqrt{-\det g} \sqrt{\det N_{ab}}\,  V[g_{\mu\nu}, \nabla_\mu \ln N_{ab},\nabla_\mu]\,,  
\end{equation}
where $V[g_{\mu\nu}, \nabla_\mu N_{ab},\nabla_\mu]$ is an scalar with scaling dimension of $[V]=D + n z$ constructed out from its argument.  Since $ N_{ab}$  is a matrix,  $V$ includes the trace operator. $V$ can be decomposed as follows
\begin{equation}
V[g_{\mu\nu}, \nabla_\mu \ln N_{ab},\nabla_\mu] = 
 \sum_{p=0}^{D+nz} V^{D+nz-p}_{i_1\cdots i_p}[g]\, f^{i_1\cdots i_p}[\nabla^i \ln N_{ab}, \nabla^i]\,, 
\end{equation}
wherein $V^{D+nz-p}_{i_1\cdots i_p}[g]$ is a tensor of rank $p$ containing $D+nz-p$ derivatives constructed out from the  space-time metric, and its covariant derivatives. $f^{i_1\cdots i_p}$ again being an scalar, is the most general polynomial constructed out from $\nabla \ln N_{ab}$ and $\nabla$ that carries the ``$i_1\cdots i_p$'' indices. Note that it is \eqref{Vgen} which governs the dynamics of the theory in the UV point. Again we choose this term such that the equations of motion for $g_{\mu\nu}$ and $N_{ab}$ are third order.

There exists another quantity that transforms nicely under \eqref{gen-foil}
\begin{equation}
K_{\mu\nu a} \,=\, \frac{\partial g_{\mu\nu}}{\partial q_a} - g_{\eta \nu} \nabla_{\mu} N^{\eta}_{a}- g_{\eta \mu} \nabla_{\nu} N^{\eta}_{a}\,.
\end{equation}
When we had only one extra q-dimension, the  part of the action that included $K$  was quadratic in term of $K$. This is not the case when more than one q-dimension exists. Suppose that the kinetic part of the action was quadratic in term of $K$, then making this part dimensionless would have required $D+nz-2z=0$   the solution of which results to negative scaling dimension for $V$ in case of $n>2$. Since we like $V$ to be polynomial of positive degree in term of derivatives of $g_{\mu\nu}$ and $N_{ab}$,  we are discard that action is quadratic in term of $K$. We further choose the `kinetic'  part of the action in the UV point to include $2n$ numbers of the $K$ tensor
\begin{equation}
S_{k} \,=\, \frac{1}{k^2} \int d^nq d^Dx \sqrt{-\det g} \sqrt{N_{ab}}\, f_1^{\mu_1\cdots \mu_{2n} \nu_1 \cdots \nu_{2n}}[g]\, f_2^{a_1\cdots a_{2n}}[N] \prod_{i=1}^{2n}K_{\mu_i\nu_i a_i}\,,
\end{equation}
 where $f_1^{\mu_1\cdots \mu_{2n} \nu_1 \cdots \nu_{2n}}[g]$ is a tensor polynomial in term of $g^{\mu\nu}$, and $f_2^{a_1\cdots a_{2n}}[N]$ is polynomial in term of $N^{ab}$ where $N^{ab}$ is the components of the inverse of $N_{ab}$. We currently do not need to present  an explicit form for $f_1$ and $f_2$. Requiring $k$ to be dimensionless then results to 
\begin{equation}\label{Sd}
D+nz -2 nz = 0 \to D = nz
\end{equation}    
The equations of motion in the IR limit by definition are second order and we have $(D+n) (D+n+1)/2$ variables. So to single out a unique solution in the IR limit we need $(D+n)(D+n+1)$ boundary conditions. As we apply our procedure to \eqref{extra-q}, we get third order equations in the UV point for each variable. In the UV point, we have $D(D+1)/2$ plus to $n(n+1)/2$ variables. So in total we need 
\begin{equation}
\frac{3}{2} (D(D+1)+n (n+1)) 
\end{equation}
`boundary conditions' to single out a unique solution in the UV point. Perhaps the dynamical consistency of the theory requires the same number of boundary conditions  in order to single out a single solution in the UV and IR point. So the consistency demands
\begin{equation}\label{ConEq}
\frac{3}{2} (D(D+1)+n (n+1))=  (D+n) (D+n+1)  
\end{equation}
For $n=1$, the above is solved by $D=1$ and $D=2$. For $n=2$, it is solved by $D=1$ and $D=6$.  Note that for $n=3,4,5$, \eqref{ConEq} does not have any solution for $D$ ($D$ should be a natural number).  For $n=6$, \eqref{ConEq} is solved by $D=2$ and $D=21$. From physical point of view, larger values for $n+D$ ( number 
of the total dimensions ) sounds not plausible.  So for 
\begin{equation}
(D,n) \in \{(1,2),(2,1),(6,2),(2,6),(6,21),(21,6)\}\,,
\end{equation}
there exists the same amount of physical information available at the UV and IR point of the theory.  Note that for each of the above choices, \eqref{Sd} gives the scaling dimension of the q-dimensions. Noticing that all the problems of original  Horava  gravity are due to  not having the same number of degrees of freedom  around the IR and UV points, we count on the possibility to have the same number of degrees freedom in IR and UV points of our realisation (beside not needing to recover full diff. symmetry in the IR limit) as an indication that our proposal will  pass further consistency check between the UV and IR point of the theory.  Knowing  how the nonlinear theory at the UV point of the proposal explicitly runs toward the IR point, however, proceeds any attempt that aims to further investigate  the dynamical consistency  between UV and IR point. The UV point, however, is linear in term of the Riemann tensor.  This might simplify the effort to find how  the UV point runs toward the IR point.

\providecommand{\href}[2]{#2}\begingroup\raggedright

\end{document}